\documentclass[pra,preprintnumbers,tightenlines,twocolumn,superscriptaddress]{revtex4-1}
\usepackage{braket}
\usepackage{array}
\usepackage{dsfont}
\usepackage{amsmath}
\usepackage{pifont}%
\usepackage{amssymb}

\usepackage{amsfonts}
\usepackage{mathtools}
\usepackage{graphicx}% Include figure files
\usepackage{dcolumn}% Align table columns on decimal point
\usepackage{bm}% bold math
\usepackage{multirow}
\usepackage{leftidx}
\usepackage{color}
\usepackage{mathtools}
\usepackage{MnSymbol}
\usepackage[mathscr]{eucal}
\usepackage[german,english]{babel}
\usepackage[capitalize]{cleveref}
\usepackage[utf8]{inputenc}

\newcommand{\E}[1][\empty]{
	\ifthenelse{\equal{#1}{\empty}}
	{\mathbb{E}}
	{\mathbb{E}\left( #1 \right)}
}
\renewcommand{\exp}[1][\empty]{
	\ifthenelse{\equal{#1}{\empty}}
	{\mathrm{exp}}
	{\mathrm{e}^{#1}}
}

\newcommand{\psit}[1][\empty]{%
	\ifthenelse{\equal{#1}{\empty}}
	{\psi_t}
	{\psi_t^{(#1)}}
}

\newcommand{\npsit}[1][\empty]{%
	\ifthenelse{\equal{#1}{\empty}}
	{\tilde\psi_t}
	{\tilde\psi_t^{(#1)}}
}

\newcommand{\be}{\begin{equation}}
\newcommand{\ee}{\end{equation}}

\usepackage{ulem}  %unter- (\uline{important}) und durchstreichen (\sout{wrong})
\newcommand{\oalex}[1]{{\color{blue}{}}}% \sout{#1}}}

\definecolor{olive}{RGB}{107,142,35}

\definecolor{orange}{RGB}{255,139,61}

\graphicspath{{{./}}}
\begin{document}
\title{Generalized local frame transformation theory for ultralong-range Rydberg molecules}

\author{P. Giannakeas}
\email{pgiannak@pks.mpg.de}
\affiliation{Max-Planck-Institut f\"ur Physik komplexer Systeme, N\"othnitzer Str.\ 38, D-01187 Dresden, Germany }
	
\author{Matthew T. Eiles}
\email{meiles@pks.mpg.de}
\affiliation{Max-Planck-Institut f\"ur Physik komplexer Systeme, N\"othnitzer Str.\ 38, D-01187 Dresden, Germany }

\author{F. Robicheaux}
\email{robichf@purdue.edu}
\affiliation{Department of Physics and Astronomy, Purdue University, 47907 West Lafayette, IN, USA}
\affiliation{ Purdue Quantum Science and Engineering Institute, Purdue University, West Lafayette, Indiana 47907, USA}

\author{Jan M. Rost}
\email{rost@pks.mpg.de}
\affiliation{Max-Planck-Institut f\"ur Physik komplexer Systeme, N\"othnitzer Str.\ 38, D-01187 Dresden, Germany }

\begin{abstract}
	A detailed theoretical framework for highly excited Rydberg molecules is developed based on the generalized local frame transformation.
	 Our approach avoids the use of pseudopotentials and yields analytical expressions for the body-frame reaction matrix. The latter is used to obtain the {molecular} potential energy curves, but equally it can be employed for photodissociation, photoionization, or other processes.
	{To illustrate the reliability and accuracy of our treatment} we consider the Rb$^*-$Rb Rydberg molecule and compare our treatment with state-of-the-art alternative approaches.
As a second application, the present formalism is used to re-analyze the vibrational spectra of Sr$^*-$Sr molecules, providing additional physical insight into their properties and a comparison of our results with corresponding measurements.
\end{abstract}
\maketitle
\section{Introduction}
Rydberg molecules constitute one of the most exotic physical systems in quantum chemistry.
They are effectively three-body systems composed of a ground state neutral atom (i.e. {\it perturber}), a positively charged core, and an electron; the subsystem formed by the latter two is a Rydberg atom.
The Rydberg electron mediates an interaction that binds together the perturber and the ionic core.
Early theoretical studies showed that the delicate nature of the binding mechanism results in the formation of a class of weakly bound molecules with bond lengths on the order of a few hundred nanometers \cite{Greene2000,hamilton_jpb_2002}.
These ultralong-range Rydberg molecules (ULRMs) are subdivided into ``trilobite'' \cite{Greene2000} and ``butterfly'' \cite{hamilton_jpb_2002} molecular species originating from the $S-$ and $P-$wave ``electron-perturber'' interactions.
Although quite fragile, both ULRMs were recently experimentally realized and observed \cite{Bendkowsky,boothtrilobite, butterfly}.
One striking attribute of the ULRMs is that -- despite the homonuclear nature of their constituents -- they can possess huge dipole moments in the range of kilodebye \cite{PfauSci,SethHosseinMolPhys}.
Also, due to the resonant $P-$wave ``electron-atom'' interaction, the butterfly molecules exhibit much deeper binding energies than the trilobite ones. The state of the art of these molecules is reviewed in Refs.~\cite{eiles_revJPB_2019a,SadeghpourReview,HamburgReview}. 

From a theoretical viewpoint, the ULRM Hamiltonian possesses a fundamental attribute that has not received explicit attention: it exhibits local symmetries in different regions of configuration space.
The exploitation of these local symmetries permits a more compact description of the physics of ULRMs and furthermore can be easily generalized to more complex scenarios, e.g. ULRMs with multiple perturbers.
One particular theoretical toolkit which can be employed in systems that possess local, rather than global, symmetries is the {\it local frame transformation theory} (LFT).

In the most general case these local symmetries can be associated with incompatible sets of approximately good quantum numbers which then are inter-related by the LFT.
U. Fano in Ref.~\cite{fano_pra_1981} and D. Harmin in Ref.~\cite{harmin_pra_1982} introduced the fundamental constituents of LFT theory in order to describe the Stark photoabsorption spectra of alkali Rydberg atoms.
The LFT concept possesses a versatile scope capable of describing a plethora of different physical systems well beyond this original application. Its applications include, for example, studies of ultracold atoms \cite{GrangerBlume2004prl,Giannakeas2012pra,Zhang2013pra, giannakeas2013prl,blume_pra_2017}, atoms in the presence of electric or magnetic fields \cite{harmin1986precise,RauWong1988pra,WongRauGreene1988pra, SlonimGreene1991,greene1987pra} or generic trapping potentials \cite{robicheaux_PRA_2015}, or Rydberg atoms with two valence electrons \cite{Armstrong1993prl,Armstrong1994pra,Robicheaux1999pra}.
In the field of molecular physics, the LFT theory provides an insightful description of processes such as electron-molecule collisions \cite{chang_PRA_1972,fabrikant_jpb_1983,clark_pra_1979}, dissociative recombination of $H_3^+$\cite{kokoulineprl2003,JungenPrattPRL}, and Stark photoabsorption spectra in molecular Hydrogen \cite{sakimoto1989jbp}.
In addition, the LFT theory has been extended to investigate the rovibrational spectra of diatomic molecules  \cite{greene1985molecular} or the electronic excitation of molecular ions \cite{priceLocalFrameTransformation2019a}.
However, for many years the LFT theory lacked a systematic pathway to improve its accuracy by including, for example, the physics of energetically strongly closed channels \cite{giannakeas_PRA_2016}.
For this reason, the \textit{generalized} local frame transformation theory (GLFT) was developed. It resolves the lack of closed channels in the original theory, providing a concrete framework based on more rigorous physical grounds.

In this study, we develop a non-perturbative theoretical framework for ULRMs which exploits the corresponding local symmetries.
Our method combines the GLFT formalism with key ideas from multichannel quantum defect theory (MQDT)\cite{seaton1983quantum}, going beyond previous studies \cite{du_JCP_1989,du_PRA_1987a}.
MQDT permits us to treat the multichannel scattering of the Rydberg electron in the polarization potential of the neutral atom in a compact manner where {\it all} the relevant physics is encapsulated in the corresponding body-frame reaction matrix $K$.
The GLFT treatment allows us to obtain $K$ analytically without using the pseudopotential theory developed by Fermi and Omont \cite{Fermi,Omont}.
In this manner, our theoretical framework avoids the limitations inherent to these methods~\cite{Fey,EilesSpin}.
For example, the GLFT approach guarantees the convergence of the potential energy curves within a particular electronic $n$ Rydberg manifold without needing to add extra ones as in the case of the diagonalization method of Omont's pseudopotential theory.

We apply the GLFT method to the calculation of the potential energy curves of both Rb$_2$ and Sr$_2$ ULRMs. In the case of rubidium, we compare the GLFT theory results with those obtained by diagonalizing the Omont pseudopotential.
Results from both methods for high electron Rydberg manifolds ($n=30$) differ quantitatively, but agree qualitatively.
However, at lower manifolds, i.e. $n=10$, the diagonalization treatment exhibits even substantial qualitative differences from the GLFT approach.
In the case of strontium, we investigate the ULRMs associated with the Rydberg $s$ state since the vibrational spectra calculated using the GLFT can be compared directly to experiment \cite{killian2015, killian2016}.
This allows us to extract more precise information about the zero energy electron-Sr $S-$wave scattering length.

 The GLFT method gives identical results as  the  Green's function treatment developed in Refs.~\cite{hamilton_jpb_2002,khuskivadze_pra_2002}. 
 However, one major advantage of the GLFT is that its scope is much more general.
The GLFT treatment encompasses all the relevant physics of the Rydberg atom interacting with the neutral one in terms of the body-frame $K-$matrix which can be utilized to describe processes of predissociation, photoabsorption, or  angular momentum changing collisions, as was pointed out also in Ref.~\cite{du_PRA_1987a}.
Moreover, linking our method with the Coulomb-Stark frame-transformation in Ref.~\cite{giannakeas_PRA_2016} permits the study of Rydberg molecules in external electric fields even in the regime where electronic $n-$manifolds are strongly mixed.
Owing to the modular structure of the GLFT framework, the treatment presented here can also be easily extended to situations involving multiple perturbers providing a compact and accurate description of the corresponding potential energy surfaces.

This work is organized as follows: \cref{sec:theory} lays out the concepts of our GLFT-MQDT theory for ULRMs.
\cref{sec:resndis} is devoted to detailed comparisons with alternative approaches and ends with a reanalysis of Sr$_2$ experimental spectra. Since the derivation in \cref{sec:theory} is quite involved, in our concluding section \cref{sec:sumncon} we provide a summary of the GLFT method and highlight the crucial expressions from the derivation before concluding. 

\section{Theoretical Method}\label{sec:theory}
\subsection{The Hamiltonian of the three-body system and general considerations}\label{sec:generalcons}
\begin{figure}[t!]
	\centering
	\includegraphics[scale=0.2]{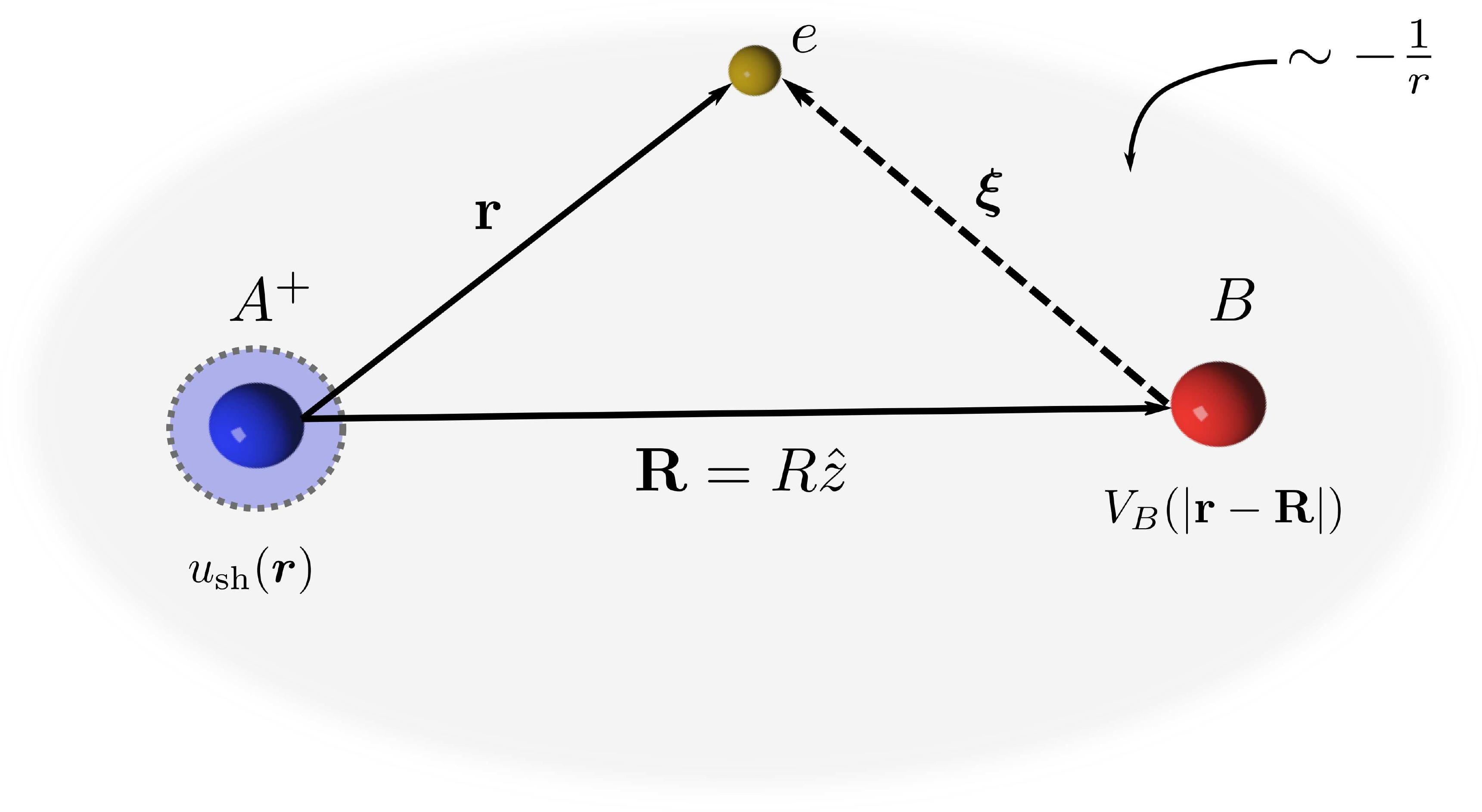}
	\caption{(color online) Schematic illustration of the few-body system of a highly excited Rydberg atom ($A^+-e$) in the presence of a neutral perturber ($B$) placed at an internuclear distance $\bm R=R \hat{z}$. The Rydberg electron $e$ at distances $\bm r$ away from the positively charged core atom $A^+$ experiences a Coulomb potential indicated by the grey shaded area. The blue shaded area depicts the range $\bm r_A$ of the residual core potential, i.e. $u_{\rm sh}(\bm r)$. $V_B(|\bm{r}-\bm{R}|)$ refers to the polarization potential between the Rydberg electron and the neutral perturber $B$.}
	\label{fig:fig1}
\end{figure}

We are interested in the ULRM system, composed of a highly excited Rydberg atom $A$ and a neutral ground state atom $B$ and shown schematically in \cref{fig:fig1}.
To handle the multichannel scattering of the Rydberg electron by the ground state atom B and to obtain a body-frame K-matrix free from singularities, we will employ ideas from the GLFT and MQDT theory.
The pair of quantum numbers $l$ and $m$ indicate the orbital angular momentum of the Rydberg electron and its projection relative to the Rydberg core $A^+$, respectively.
Similarly, the quantum numbers $L$ and $M$ denote the electron's orbital and azimuthal angular momentum relative to the perturber $B$, respectively. 
The internuclear axis $\vec R = R\hat z$ is aligned with the $z$ axis, and thus the azimuthal quantum numbers  $m$ and $M$ are conserved, i.e. $m=M$.
 The position of the Rydberg electron is $\bm r$ with respect to the Rydberg core, and $\bm \xi= \bm r - \bm R$ with respect to the perturber (see \cref{fig:fig1}).
The full Hamiltonian in the frame of reference of the Rydberg core reads
\begin{equation}
	H=-\frac{1}{2}\nabla_r^2+u_{\rm sh}(\bm r)-\frac{1}{r}+V_B(|\bm{r}-\bm{R}|).
	\label{eq:hamiltonian}
\end{equation}
The term $ u_\text{sh}(\bm r)$ is a parameterization for the interaction between the Rydberg electron and the many electrons of the residual atomic core. This interaction is short-range, i.e. it vanishes for distances $r>|\bf r_A|$, and so its effect on the electronic motion is encapsulated by a set of quantum defects $\mu_l$.
Note that in \cref{eq:hamiltonian} and throughout this study we use atomic units unless otherwise specified.

At distances $\bm r\approx \bm R$, the Rydberg electron interacts with the perturber via a short-range potential, $\hat V_B$, of asymptotic form
\begin{equation}
	\label{eq:polarizationinteraction}
	V_B(\xi) \sim -\frac{\alpha}{2\xi^4},\quad {\rm{with}}~{\xi}=|\boldsymbol{r}-\boldsymbol{R}|,
\end{equation}
where $\alpha$ is the static polarizability of the neutral atom. At smaller $\xi$ this interaction also includes electron correlation, interaction with the perturber's core electrons, and exchange interaction. As with the quantum defects, the effect of $V_B$ is imprinted on the wave function at large distance by a set of phase shifts. 

Since the interactions between the electron and the Rydberg core and perturber are limited to small volumes around each site, over the rest of the Rydberg's orbit the electron only experiences the Coulomb potential.
In the spirit of MQDT, we postpone imposition of the {\it physical boundary conditions} to a later point in our derivation and  impose ``standing-wave'' boundary conditions for now.
This simplifies tremendously the analytic manipulations while keeping the scope of the treatment as general as possible.
The Rydberg electron's wave function at $r\to\infty$ reads
\begin{equation}
	\label{eq:asymptoticbc}
	\Phi_{lm}(\bm r,R) = \sum_{l'}Y_{l'm}(\hat r)\left[ f_{l'}^c(r) \tilde \delta_{l'l} -  g_{l'}^c(r)K_{l'l}(R)\right],
\end{equation}
where $ f_l^c(r)$ and $ g_l^c(r)$ are {\it energy normalized} regular and irregular Coulomb functions, respectively.
Note that the symbol $\tilde \delta_{l'l}$ denotes Kronecker's delta.
The pair $( f_l^c, g_l^c)$ are evaluated at an electronic energy $\epsilon(R)$ which depends on the internuclear separation $R$, and are defined with the origin centered at the ionic core.
$K_{l'l}(R)$ are the elements of the real, symmetric, body-frame $K$-matrix and satisfy
\begin{equation}
	\label{eq:Kmat}
	K_{l'l}(R)=-\pi\braket{f_{l'}^c|V_B|\Phi_{lm}(R)}.
\end{equation}
Note that the preceding equation is an exact relation given by the Lippmann-Schwinger equation that includes $u_{sh}(\bf{r})$, $V_B(\bf{r})$, and the Coulomb potential.

Once the $K$-matrix is determined, we can impose the proper physical asymptotic boundary conditions by requiring that the electronic wave function vanishes at large $r$. 
This condition provides us with the bound states of the Rydberg electron in the presence of the perturber, yielding the following determinantal equation:
\begin{equation}
	\label{eq:determinantal1}
	\det\left[\tilde \delta_{l'l}+\cot\pi\nu K_{l'l}(R)\right]=0,
\end{equation}
where the principal quantum number $\nu =1/\sqrt{-2\epsilon(R)}$ is determined by the discrete electronic energy $\epsilon(R)$. These energies, parametrically depending on the internuclear separation, define the set of Born-Oppenheimer potential energy curves. The key quantity determining these potential curves is evidently the $K$-matrix.
Therefore, the following subsections show how to analytically obtain the $K$-matrix using the GLFT.

\subsection{Quantum-defect-shifted Coulomb functions and smooth $\overline{K}$-matrix}
In the same spirit as Refs.~\cite{du_PRA_1987a,du_JCP_1989} it is desirable to express the  electronic wave function of \cref{eq:asymptoticbc} using an alternative pair of energy-normalized regular and irregular Coulomb functions, namely the so-called {\it quantum-defect-shifted (QDS)} Coulomb functions.
This alternative pair of Coulomb solutions permits us to eliminate the potential term $u_{\rm sh}(\bm r)$ from \cref{eq:hamiltonian} since its effects are encapsulated in the atomic quantum defects $\mu_l$.
The energy-normalized QDS pair of solutions are related to the conventional regular and irregular Coulomb functions ($f_l^c,~g_l^c$) according to
\begin{subequations}
	\begin{align}
		\label{eq:coulombfunctions2}
		F_l(r) =  f_l^c(r)\cos\pi\mu_l -  g_l^c(r)\sin\pi\mu_l,\\
		G_l(r) =   f_l^c(r)\sin\pi\mu_l + g_l^c(r)\cos\pi\mu_l, \label{eq:coulombfunctions2G}
	\end{align}
\end{subequations}
The QDS pair of solutions $(F_l,~G_l)$ are associated with the $\overline{K}$-matrix,  {\it not} the $K$-matrix of \cref{eq:asymptoticbc}. These matrices are inter-related via a simple matrix transformation (see discussion in \cref{sec:elimin}).

The electronic wavefunction from \cref{eq:asymptoticbc} at large distances is expressed as a linear combination of the QDS Coulomb functions,
\begin{equation}
	\label{eq:qdswfn}
	\Psi_{lm}(\bm r, R)=\sum_l Y_{lm}(\hat r) [F_{l'}( r) \tilde \delta_{l'l}-G_{l'}( r)	\overline{K}_{l'l}(R)],
\end{equation}
and in general it fulfills the Lippmann-Schwinger equation
\begin{equation}
	\label{eq:lippmanschwinger}
	\ket{\Psi_{lm}(R)} = \ket{{F}_{l}(R)} + \hat {G}_C^{\rm{QDS}}\hat V_B\ket{\Psi_{lm}(R)},
\end{equation}
where $\hat{G}_C^{\rm{QDS}}\equiv[\epsilon(R)+\nabla_r^2/2+1/r-u_{\rm{sh}}(r)]^{-1}$ is the quantum-defect-shifted {\it principal-value} Coulomb Green's function and $V_B$ is the electron-perturber interaction potential discussed above.
This Green's function reads in terms of the QDS Coulomb functions as
\begin{equation}
	\label{eq:qdcoulombgreen}
	G_C^\text{QDS}(\bm r,\bm r') = \pi\sum_{lm}Y_{lm}^*(\hat r)F_l(r_<)G_l(r_>)Y_{lm}(\hat r'),
\end{equation}
where $r_< = \text{min}(r,r')$ and $r_> = \text{max}(r,r')$. Note that $G_C^\text{QDS}(\bm r,\bm r')$ obeys {\it standing wave} boundary conditions at large distances.
The Lippmann-Schwinger equation in \cref{eq:lippmanschwinger} provides us with the $\overline K$-matrix, $\overline K_{l'l}(R)=-\pi\braket{F_{l'}(R)|V_B|\Psi_{lm}(R)}$.
As was shown in Refs. \cite{robicheaux_PRA_2015,giannakeas_PRA_2016}, by employing the Schwinger identity  the $\overline{K}$-matrix can be expressed in the following symmetric form:

\begin{align}
	\label{eq:schwidentity}
	&~&\overline K_{l'l}(R)=-\pi\braket{F_{l'}(R)|\hat V_B \hat M^{-1}\hat V_B|F_l(R)}~~\\
	&~&\rm{with}~ ~ \hat M = \hat V_B-\hat V_B \hat{G}_C^{\text{QDS}}\hat V_B.\nonumber
\end{align}

Similar to Refs.\cite{robicheaux_PRA_2015,giannakeas_PRA_2016}, the $\overline{K}$ matrix elements in \cref{eq:schwidentity} can be expanded over a complete basis set of wave functions $\ket{\phi_{LM}}$ which are solutions of the electron-perturber ($e-B$) subsystem.
A key assumption, justified by the short-range nature of the $e-B$ interactions, is that over the small region of space around the perturber where the polarization potential $\hat V_B$ prevails, the Coulomb field is approximately constant.
The $\ket{\phi_{LM}}$ wavefunctions fulfill the following Schr\"odinger equation:
\begin{equation}
	\label{eq:eBse}
	\bigg[-\frac{\nabla_\xi^2}{2}+ V_B(\bm \xi)\bigg]\phi_{LM}(\bm \xi)=\left(\epsilon(R) + \frac{1}{R}\right)\phi_{LM}(\bm \xi).
\end{equation}
At distances $\xi$ larger than the range of the $\hat V_B$ potential, $\phi_{LM } (\bm \xi)$ acquires its asymptotic form 

\begin{equation}
	\label{eq:Basymptoticbc}
	\phi_{LM}(\bm \xi) = Y_{LM}(\hat \xi)\left[f^0_{LM}(\xi) - g^0_{LM}(\xi)\tan\delta_L\right],
\end{equation}
where $(f^0_{LM}(\xi),g^0_{LM}(\xi))$ are the energy-normalized regular and irregular field-free solutions, namely the spherical Bessel and Neumann functions, respectively. 
The term $\delta_L\equiv\delta_L(k)$ is the energy-dependent scattering phase shift induced by the polarization potential $\hat V_B$.
 The relative momentum of the $e-B$ subsystem is  $k\equiv k(R)$, and its $R$-dependence comes from the relation $k(R)=\sqrt{2\epsilon(R)+2/R}$.
 The quantum numbers $L$ and $M$ label the angular and azimuthal momentum {\it with respect to the perturber B} and they are different from the quantum numbers $l$ and $m$ which are associated with wave functions centered at the ionic core $A^+$.
Also the azimuthal quantum numbers $m$ and $M$ are conserved; thus they remain the same for a wavefunction that is either centered at $A^+$ or $B$.
The $\overline{K}$-matrix in \cref{eq:schwidentity} expanded over the states $\ket{\Phi_{LM}}$ takes the form
\begin{subequations}
	\begin{align}
		&\overline{K}_{l'l}(R) = -\pi\sum_{LL'}\mathcal{C}_{l'm,LM}(R)[M]^{-1}_{LL'}\mathcal{C}_{L'M,lm}^T(R), \label{eq:kmatrix1}\\
		&{\rm{with}}~~\mathcal{C}_{L'M,lm}^T(R)=\braket{\phi_{L'M}|\hat V_B|F_{lm}(R)}~\rm{and} \label{eq:kmatrix1b}\\
		&	M_{L'L} = \bra{\phi_{L'M}}\hat V_B - \hat V_B \hat G_C^{\text{QDS}}\hat V_B\ket{\phi_{LM}}. \label{eq:kmatrix1c}
	\end{align}
\end{subequations}

Evidently, the computation of the $\overline{K}$-matrix involves the evaluation of the matrix elements in \cref{eq:kmatrix1b,eq:kmatrix1c}.
However, the terms $\mathcal{C}_{L'M,lm}^T(R)$ and $M_{L'L}$ require the computation of complicated volume integrals which involve functions that are centered at two different locations of the configuration space: recall that the QDS function $F_{lm}(\bm r;R)$ is centered around $A^+$ while the wave function $\phi_{LM}(\bm \xi )$ is centered around $B$.
This challenge can be efficiently avoided by introducing the core idea of the GLFT approach, which enables us to transform the complicated volume integrals into simpler surface ones.
In the following subsection we will introduce and derive the local frame transformation for Rydberg molecules, which inter-relates the functions $\ket{F_{l'm}(R)}$ with the solutions $\ket{f^0_{LM}}$. 
Such a transformation is valid only in a {\it local }region around $B$, where the motion of the Rydberg electron is mainly influenced by the polarization potential.

\subsection{Local frame transformation for two different scattering centers}
According to \cref{fig:fig1} the three-body system, $A^+-e-B$, can be divided into two subsystems: (i) the Rydberg atom ($A^+-e$), which is conveniently described by the QDS Coulomb functions centered on the Rydberg core, and (ii) the electron-perturber ($e-B$) complex that is addressed by free particle scattering solutions relative to the perturber B which is located at internuclear distances $R$ from the Rydberg core.
In the following, we derive the relevant expressions which connect the regular QDS Coulomb function $\ket{F_{l'm}(R)}$ centered at $A^+$ with the energy normalized regular field-free solutions $\ket{f^0_{LM}}$ centered at $B$.

As a first step, at distances $\bm r \approx \bm R$ we express the regular QDS Coulomb functions as a linear combination of the two linearly independent field-free regular and irregular functions that are centered at the Rydberg core $A^+$:

\begin{align}
	F_{lm}(\bm r)=&\frac{\pi R^2}{2}\bigg[f^0_{lm}(\bm r) \mathcal{W}\{F_{lm},g^0_{lm}\}_R\nonumber \\
	&-g^0_{lm}(\bm r) \mathcal{W}\{F_{lm},f^0_{lm}\}_R\bigg]~ ~{\rm{for}}~\bm r\approx \bm R,
	\label{eq:expqdsF}
\end{align}
where the pair functions $(f^0_{lm},g^0_{lm})$ are the energy-normalized spherical Bessel and Neumann functions respectively.
The term $\mathcal{W}\{\cdot,\cdot\}_R$ indicates the Wronskian evaluated at $R$. 
It should be noted that \cref{eq:expqdsF} only applies in the vicinity around the perturber $B$.
The prefactor $\pi R^2/2$ is the inverse of the Wronskian of the regular and irregular field-free solutions at $r=R$, i.e. $\mathcal{W}\{f^0_{lm},g^0_{lm}\}_R=2/(\pi R^2)$.

The right-hand side of \cref{eq:expqdsF} is still expressed in terms of field-free regular and irregular functions relative to $A^+$.
These can be analytically re-expanded into field-free regular functions centered at a different point by exploiting basic attributes of bispherical harmonics [see Eq. (34) in Ref. \cite{varshalovich_angmom}].
This identity relates the spherical functions, i.e. $f_{lm}^0(\bm r)$ and $g_{lm}^0(\bm r)$, to \textit{spherical Bessel functions relative to $B$, i.e. $f_{LM}^0(\bm\xi)$} with $\bm \xi = \bm r -\bm R$, via
\begin{subequations}
	\begin{align}
		\begin{Bmatrix} f^0_{lm}(\bm r) \\ g^0_{lm}(\bm r) \end{Bmatrix}
		&=
		\sum_{LM}f^0_{LM}(\bm \xi)
		\begin{Bmatrix}
			\mathcal{J}_{LM,lm}^T(R)\\
			\mathcal{N}_{LM,lm}^T(R)\\
		\end{Bmatrix} \label{eq:varseq1}
		\\
		\begin{Bmatrix} \mathcal{J}_{LM,lm}^T(R) \\ \mathcal{N}_{LM,lm}^T(R) \end{Bmatrix}
		&=
		\sqrt{\frac{2L+1}{2l+1}}\sum\limits_{\beta=0}^\infty i^{L + \beta -l}(2\beta + 1)\times\nonumber \\
		&\times C_{L0\beta0}^{l0}C_{LM\beta 0}^{lm}
		\begin{Bmatrix}
			j_\beta(kR)\\
			n_\beta(kR)\\
		\end{Bmatrix},\label{eq:varseq2}
	\end{align}
\end{subequations}
where $C_{j_1m_1,j_2m_2}^{j_3m_3}$ are the Clebsch-Gordan coefficients which account for the re-coupling of angular momenta relative to the Rydberg core with the corresponding angular quantum numbers that are associated with the perturber.

Substitution of \cref{eq:varseq1,eq:varseq2} into \cref{eq:expqdsF} results in the {\it local frame transformation} between the QDS regular Coulomb function and the field-free regular functions,
\begin{align}
	F_{lm}(\bm r)=\sum_{LM}f^0_{LM}(\bm \xi)\mathcal{U}^T_{LM,lm}(R;\mu_l),~ ~{\rm{with}} 
	\label{eq:lft}\\
	\begin{aligned}
		\mathcal{U}^T_{LM,lm}(R;\mu_l)=\frac{\pi R^2}{2}&\bigg[\mathcal{J}_{LM,lm}^T(R) \mathcal{W}\{F_{lm},g^0_{lm}\}_R\nonumber\\
		&-\mathcal{N}_{LM,lm}^T(R) \mathcal{W}\{F_{lm},f^0_{lm}\}_R\bigg].\nonumber 
	\end{aligned}
\end{align}
Note that, when the quantum defects $\mu_l$ vanish, \cref{eq:lft} provides us with the transformation of the regular Coulomb functions $f^c_{lm}(\bm r)$ relative to $A^+$ in terms of the field-free regular solutions $f^0_{LM}(\bm \xi)$ relative to $B$.

In addition, a similar local frame transformation can be derived for the irregular QDS Coulomb function defined in \cref{eq:coulombfunctions2G} if needed.
More specifically, as is discussed in the following subsection, the evaluation of the $M_{L'L}$ matrix elements is greatly simplified by frame transforming the Whittaker Coulomb functions which  exponentially decay at large distances.
The energy-normalized Whittaker Coulomb functions obey the relation $w^c_{lm}(\bm r)=Y_{lm}(\hat r)[g_{l}^c(r)+\cot \pi\nu f_{l}^c(r)]$ and the corresponding local frame transformation reads:
\begin{align}
	w^c_{lm}(\bm r)=\sum_{LM}f^0_{LM}(\bm \xi)\mathcal{V}^T_{LM,lm}(R),~ ~{\rm{with}} \label{eq:whitlft}\\
	\begin{aligned}
		\mathcal{V}^T_{LM,lm}(R)=\frac{\pi R^2}{2}&\bigg[\mathcal{J}_{LM,lm}^T(R) \mathcal{W}\{w^c_{lm},g^0_{lm}\}_R\nonumber\\
		&-\mathcal{N}_{LM,lm}^T(R) \mathcal{W}\{w^c_{lm},f^0_{lm}\}_R\bigg].\nonumber 
	\end{aligned}
\end{align}
With this local frame transformation in hand, we compute in the following subsection the matrix elements necessary to determine the $\overline{K}$-matrix.

\subsection{Evaluation of the $\overline{K}$-matrix in terms of the local frame transformation}
\subsubsection{The $\mathcal{C}^T_{LM,lm}(R)$ matrix elements}
The corresponding volume  integrals in the $\mathcal{C}^T_{LM,lm}(R)$ terms of \cref{eq:kmatrix1b} contain the short-range polarization potential $\hat V_B$.
As was shown in Ref.\cite{robicheaux_PRA_2015}, such integrals can generally be recast into surface ones. Given this, \cref{eq:kmatrix1b} reads
\begin{equation}
	\begin{aligned}
		\mathcal{C}^T_{LM,lm}(R)=\frac{\xi^2}{2}\int&\bigg[\phi^*_{LM}(\bm\xi)\partial_\xi F_{lm}(\bm \xi +\bm R)\\
		&-\partial_\xi \phi^*_{LM}(\bm\xi)F_{lm}(\bm \xi +\bm R)\bigg] d\Omega_\xi, 
	\end{aligned}
	\label{eq:ctermwronsk}
\end{equation}
where $d\Omega_\xi=\sin \theta_\xi d\theta_\xi d\varphi_\xi$ is the solid angle differential element around the location of the perturber B.
By employing the local frame transformation (see \cref{eq:lft}), the $\mathcal{C}^T_{LM,lm}(R)$ matrix elements become
\begin{equation}
	\mathcal{C}^T_{LM,lm}(R)=-\frac{\tan \delta_{L}}{\pi}\mathcal{U}^T_{LM,lm}(R;\mu_l).
	\label{eq:cterms}
\end{equation}

\subsubsection{The $M_{L'L}$ matrix elements}
The evaluation of the $M_{L'L}$ matrix elements in \cref{eq:kmatrix1c} poses more challenges than the $\mathcal{C}^T_{LM,lm}(R)$ terms. 
The main source of difficulties arise from the QDS Coulomb Green's function which possesses a divergence for $\bm r\to\bm r'$. 
However, in the spirit of the GLFT approach this type of divergence can be eliminated simply by adding and subtracting the field-free principal-value Green's function $\hat G_0=[k^2(R)/2+\nabla^2/2]^{-1}$ which exhibits the same pole in the limit $\bm r\to \bm r'$ \cite{giannakeas_PRA_2016}.
By forming the difference of principal-value Green's functions, i.e. $\Delta \hat G\equiv \hat G_0-\hat{G}_C^{\rm{QDS}}$, the matrix elements $M_{L'L}$ are found to obey the following relation:
\begin{align}
	M_{L'L} = -\frac{1}{\pi}\tan\delta_{L'} \tilde \delta_{L'L} +\bra{\phi_{L'M}}\hat V_B\Delta \hat G\hat V_B\ket{\phi_{LM}}.
	\label{eq:mdeltaG}
\end{align}

In order to evaluate \cref{eq:mdeltaG} it is desirable to isolate those terms in $\Delta \hat G$ which depend on the atomic quantum defects $\mu_l$.
This is achieved by expressing $\Delta \hat G$ in terms of the Coulomb Green's function $\hat G_C$ which obeys exponentially decaying boundary conditions asymptotically, i.e. {\it the physical Coulomb Green's function}. This gives
\begin{align}
	\Delta G(\bm r,\bm r')=&\bigg[G_0(\bm r,\bm r')-G_C(\bm r,\bm r')\bigg] - \Lambda(\bm r,\bm r')\nonumber\\
&+\pi\cot\pi\nu\sum_{lm}f_{lm}^c(\bm r_<) f_{lm}^c(\bm r'_>) 
	\label{eq:deltaG0Gc}
\end{align}
where
\begin{subequations}
	\begin{align}
		&G_C(\bm r,\bm r')=\pi\sum_{lm}f^c_{lm}(\bm r_<)w^c_{lm}(\bm r'_>) ~ ~{\rm and}\label{eq:coulgf}\\
		&\Lambda(\bm r,\bm r') = \pi\sum_{lm}^{l_N}\left[F_{lm}(\bm r_<)G_{lm}(\bm r_>) - f_{lm}^c(\bm r_<)g_{lm}^c(\bm r_>)\right].\label{eq:lambdaterm}
	\end{align}
\end{subequations}
All the functions that depend on the atomic quantum defects $\mu_l$ are isolated in the term $\Lambda$, which is a finite sum which terminates at  $l_N$. For $l>l_N$ all quantum defects $\mu_l$ vanish.

Using \cref{eq:deltaG0Gc} and the local frame transformation from \cref{eq:lft}, we recast the volume integrals into surface integrals as  prescribed in \cref{eq:ctermwronsk} to obtain
\begin{align}
	M_{L'L}&=-\frac{\tan \delta_{L'}}{\pi}\tilde \delta_{L'L}+\braket{\phi_{L'M}|\hat V_B (\hat G_0-\hat G_c)\hat V_B|\phi_{LM}}\nonumber\\
	&-\braket{\phi_{L'M}|\hat V_B \hat \Lambda \hat V_B|\phi_{LM}}\nonumber\\
	&+\frac{\tan \delta_{L'}\tan \delta_L}{\pi}\cot \pi \nu\sum_{lm}\mathcal{U}^T_{L'M,lm}(R;0)\mathcal{U}_{lm,LM}(R;0)\label{eq:mmatlft}.
\end{align}
The terms $\braket{\phi_{L'M}|\hat V_B (\hat G_0-\hat G_c)\hat V_B|\phi_{LM}}$ and $\braket{\phi_{L'M}|\hat V_B \hat \Lambda \hat V_B|\phi_{LM}}$, expressed in terms of  \cref{eq:varseq1}, \cref{eq:lft} and \cref{eq:whitlft}, read
\begin{widetext}
	\begin{align}
		&\bra{\phi_{L'M}}\hat V_B(\hat G_0 - \hat G_C)\hat V_B\ket{\phi_{LM}}=\frac{\tan\delta_{L'}\tan\delta_{L}}{\pi}\sum_{lm}\mathcal{J}^T_{L'M,lm}(R)\mathcal{N}_{lm,LM}(R) - \mathcal{U}^T_{L'M,lm}(R,0)\mathcal{V}_{lm,LM}(R)\label{eq:g0gcterm}\\
		&\bra{\phi_{L'M}}\hat V_B\hat \Lambda \hat V_B\ket{\phi_{LM}}=\frac{\tan\delta_{L'}\tan\delta_{L}}{\pi}\Bigg\{\sum_{lm}^{l_N}\Bigg[-\frac{1+\cos\pi\mu_l}{\sin\pi\mu_l}\left(\mathcal{U}^T_{L'M,lm}(R,\mu_l)\mathcal{U}_{lm,LM}(R,\mu_l) + \mathcal{U}^T_{L'M,lm}(R,0)\mathcal{U}_{lm,LM}(R,0)\right)\nonumber\\
		&\qquad\qquad\qquad\qquad\qquad\qquad+\frac{1}{\sin\pi\mu_l}\left[\mathcal{U}^T_{L'M,lm}(R,\mu_l)+\mathcal{U}^T_{L'M,lm}(R,0)\right]\left[\mathcal{U}_{lm,LM}(R,\mu_l)+\mathcal{U}_{lm,LM}(R,0)\right]\Bigg]\Bigg\}.\label{eq:lambdalft}
	\end{align}
\end{widetext}
From these closed-form expressions we can obtain the $\overline{K}$-matrix free from any unphysical divergences.
Since the physical boundary conditions are not imposed yet, the matrix elements of $\overline{K}$ are known for any angular momentum $l$.
This is the focus of the following subsection, which addresses the importance of the {\it energetically strongly closed channels}, i.e. $l>\nu$.

\subsection{Pre-elimination of strongly closed channels in $\overline{K}$-matrix and the connection relation to $K-$matrix}\label{sec:elimin}
In general, the motion of the Rydberg electron in the presence of a charged core and a neutral perturber can be viewed in terms of half collisions, as in the photoionization of Rydberg atoms.
From this viewpoint, the $(l,m)$ quantum numbers that are relative to the Rydberg core can be used to label the different collisional channels.
The perturber couples only the different $l$-channels since $m$ is a good quantum number, resulting in a non-diagonal $\overline{K}$-matrix.

The different $l-$channels separate into two categories: (i) {\it Weakly open channels}, which refer to $l$-states with a weak centrifugal barrier yielding a classically allowed region between the Rydberg core and the perturber and
(ii) {\it Strongly closed channels}, indicated by $l-$states associated with strong centrifugal forces imposing a classical forbidden region between the core and the perturber.
In particular, we consider all  $l\le n^*-1$ momenta in the weakly open channels, where $n^*$ is the integer lying in the interval $\nu<n^*\le\nu+1$.
All angular momenta with $l>n^*-1$ are regarded as strongly closed channels.

For these strongly closed channels, the classically forbidden region causes the corresponding regular and irregular Coulomb functions to possess imaginary energy-normalization constants; the electron's energy is below the combined centrifugal and Coulomb potential over the entire configuration space.
Since we have not yet imposed the physical boundary conditions on these wave functions, the $\overline{K}$-matrix contains those unphysical channels.

In order to properly treat the physics of the strongly closed channels, the wave function in \cref{eq:qdswfn} is partitioned into open ({\it 'o'}) and closed ({\it 'c'}) channels, as is usually done in multi-channel quantum defect theory.
Subsequently, the closed channel part of the wave function is eliminated by imposing the physical boundary conditions asymptotically, i.e. by forcing the corresponding part of the wave function  to exponentially decay at large distances.
This pre-elimination of the strongly closed channels results in a wave function that involves only the weakly open channels and is associated with a real and symmetric {\it physical} $\overline{K}$-matrix.  In compact form the $\overline{K}$-matrix reads:
\begin{equation}
	\overline{\bf K}^{\rm phys}_{oo}(R)=\overline{\bf K}_{oo}(R)-\overline{\bf K}_{oc}(R)\big[\tan \pi {\boldsymbol \nu}+\overline{\bf K}_{cc}(R)\big]^{-1}\overline{\bf K}_{co}(R),
	\label{eq:physkbar}
\end{equation}
where the collective index $o$ ($c$ ) indicates all the angular momentum that fulfill the relation $0\le l \le n^*-1$ ($l>n^*-1$ ) for $\nu<n^*\le\nu+1$ where $n^*$ is an integer. 
In total there are $N_c$ closed channels.
The matrix $\tan\pi{\boldsymbol \nu}$ is diagonal with $\nu$ defined after \cref{eq:determinantal1}.
This elimination of strongly closed channels leads to the wave function
\begin{equation}
	\Psi_{lm}(\bm r, R)=\sum_{l'}^{n^*-1} Y_{l'm}(\hat r)[F_{l'}(r)-G_{l'}(r)\overline{K}^{\rm{phys}}_{l'l}(R)].
	\label{eq:physpsi}
\end{equation}

As a final step, we compute the $K$-matrix using the $\overline{K}$-matrix by linearly combining the solutions in \cref{eq:physpsi} to construct the wave function that involves the regular and irregular Coulomb functions.
This transformation can be expressed in a compact form:
\begin{equation}
	\label{eq:kkbar}	
	\bf{K}=\left[\sin\pi\boldsymbol{\mu}+\cos\pi\boldsymbol{\mu}\overline{\bf{K}}^{\rm{phys}}\right]\left[\cos\pi\boldsymbol{\mu} - \sin\pi\boldsymbol{\mu}\overline{\bf{K}}^{\rm{phys}}\right]^{-1}, 
\end{equation}
where $\cos \pi \boldsymbol{\mu}$ and $\sin\pi\boldsymbol{\mu}$ are diagonal matrices.
Recall that $\mu_l$ indicate the atomic quantum defects.

\begin{figure}[t!]
	\centering
	\includegraphics[scale=0.35]{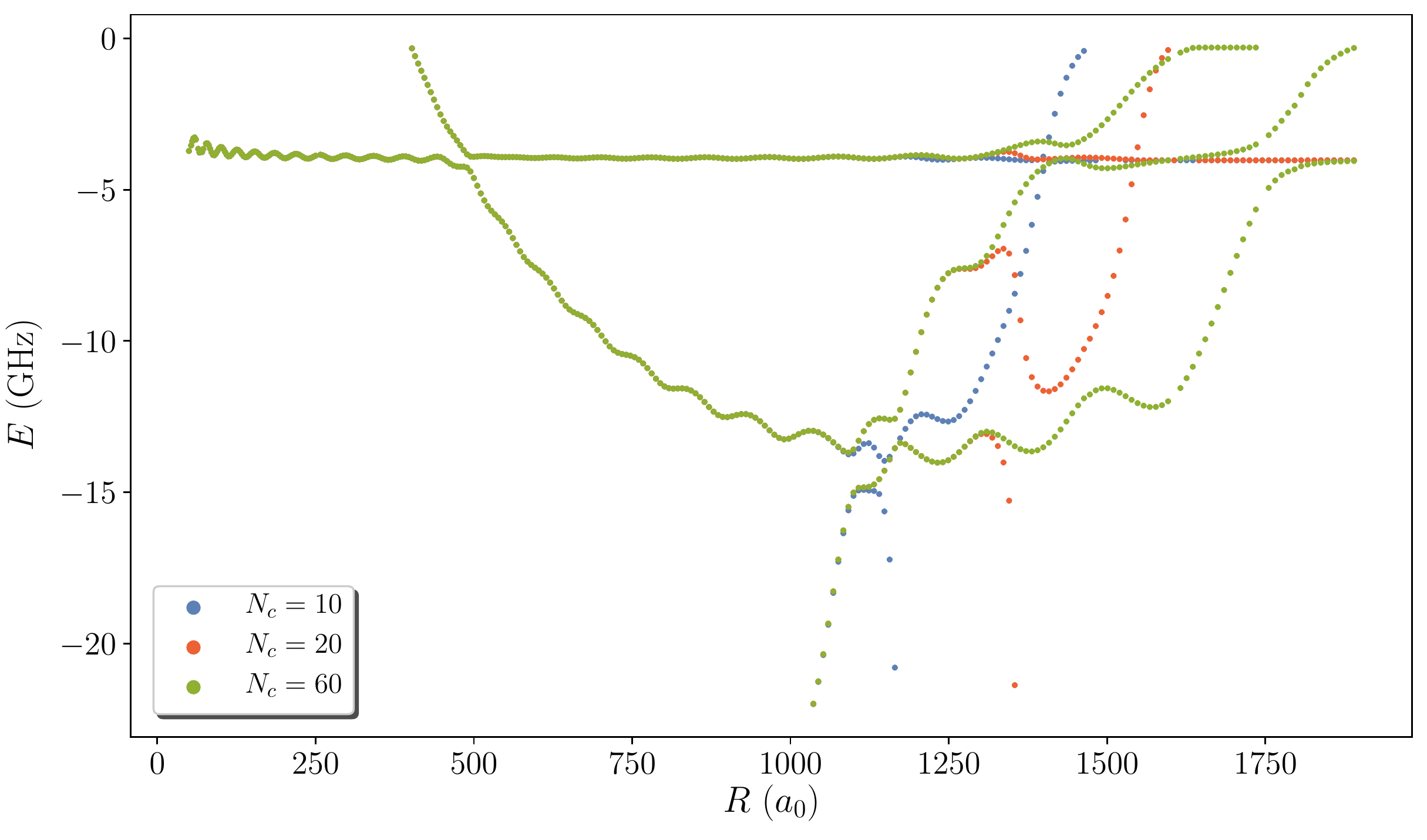}
	\caption{(color online) The $\Sigma-$molecular potential curves for the system Rb$^*-$Rb at $n=30$ computed using different numbers of strongly closed channels, $N_c$: $N_c=10$ (blue), $N_c=20$ (red) and $N_c =60$ (green).}
	\label{fig:figs-fig2-eps}
\end{figure}
Now that we have obtained a closed form expression for the physical $K$-matrix we can analyze the role of the strongly closed channels. 
Their importance is illustrated in \cref{fig:figs-fig2-eps}.
The $\Sigma$ potential energy curves for Rb$^*-$Rb near the $n=30$ degenerate Rydberg manifold are shown for three different values of $N_c$, $N_c = 10$ (blue), $N_c = 20$ (red), and $N_c = 60$ (green).
Note that the green dots represent the converged potential curve.
\cref{fig:figs-fig2-eps} illustrates that at internuclear distances $R<1100~a_0$ the trilobite potential curve is sufficiently converged by including a small number of strongly closed channels in the $\overline{K}^{\rm{phys}}$ matrix elements.
However, for $R\ge1100~{\rm{a}_0}$, the convergence depends strongly on $N_c$.
Several numerical calculations for different electronic manifolds $n$ show that the minimum number of strongly closed channels is $N_c\approx 2 n$ to ensure convergence of $\overline{K}^{\rm{phys}}$ out to and even beyond the classical turning point, $R\approx2n^2$, yielding potential energy curves converged up to 8 significant digits.
In closing, we note that the theory of Refs.~\cite{du_JCP_1989,du_PRA_1987a} neglects the strongly closed channels and therefore would not give accurate results for {\it long-range} Rydberg molecules.
Also, the present method is computationally efficient since the matrix inversion in \cref{eq:physkbar} is performed analytically by exploiting the fact that $\bar{\boldsymbol{K}}_{cc}(R)$ is a rank one matrix.
Thus the numerical solution of the transcendental equation involves a $K-$matrix whose dimensionality is $n-\rm{by}-n$ for the $n$ Rydberg manifold.

\section{Results and discussion}\label{sec:resndis}
\subsection{Rubidium}\label{sec:resndisRb}
In the following the GLFT approach is applied to the Rb$^*-$Rb system to benchmark it against other state-of-the-art methods, namely the diagonalization of Omont/Fermi pseudopotentials and the Coulomb Green's function method.
The $S-$ and $P-$wave phase shifts of the ``electron-perturber'' subsystem are obtained by a non-relativistic two electron R-matrix method \cite{EilesHetero,TaranaCurik}.
Recall that the $S- $ and $P-$wave electron-perturber phase shifts give rise to the trilobite and butterfly molecules, respectively.
The atomic quantum defects and the ``electron-perturber'' phase shifts are the only auxiliary input parameters.

\begin{figure}[t!]
	\centering
	\includegraphics[scale=0.45]{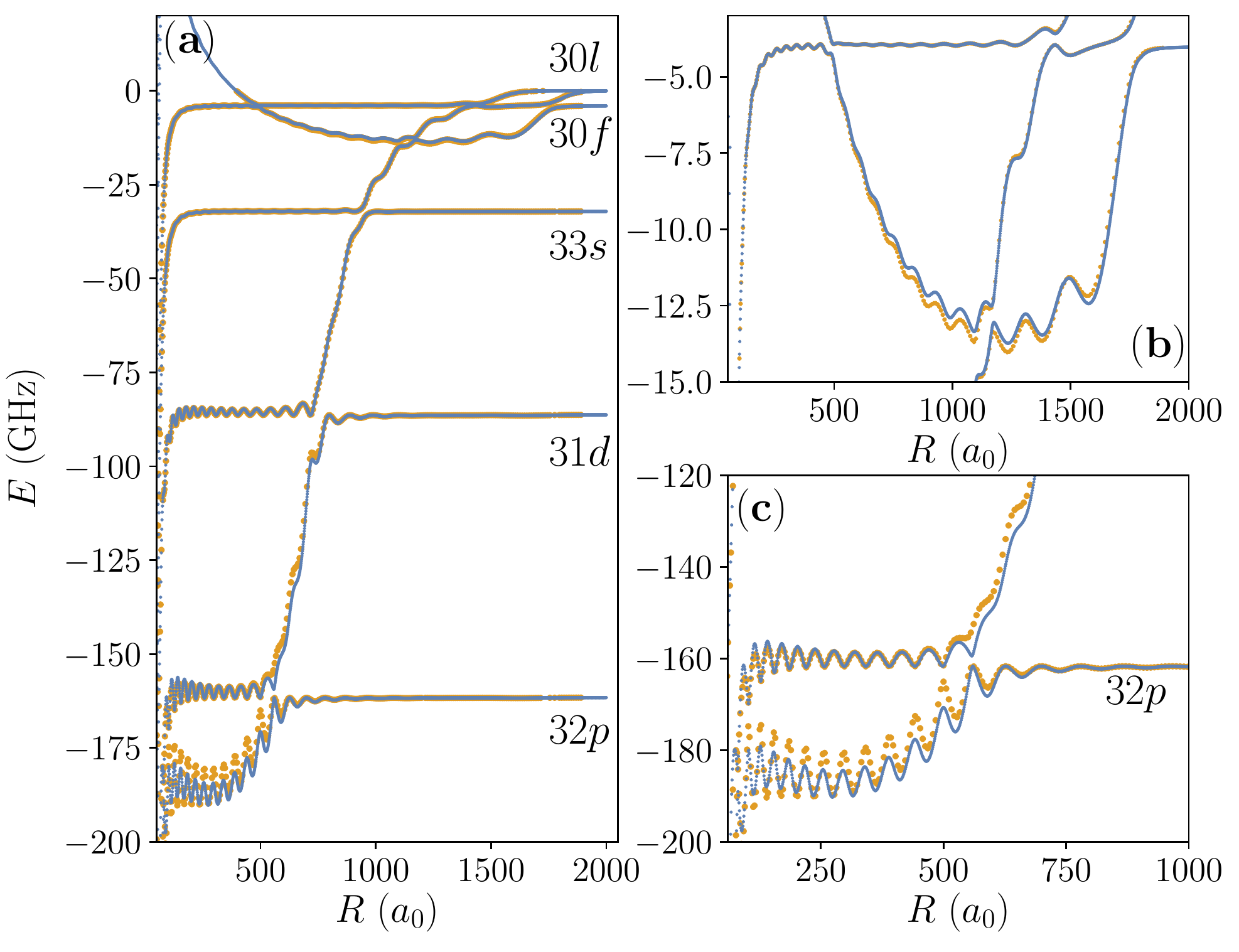}
	\caption{(color online) (a) Comparison of the $\Sigma-$molecular potentials curves obtained via the diagonalization approach (solid blue line) and the GLFT method (orange dots) for Rb$^*-$Rb at $n=30$. Panels (b) and (c) are zoom in plots of panel (a) near the trilobite and butterfly molecular curves, respectively. }
	\label{fig:figs-fig3-eps}
\end{figure}

\cref{fig:figs-fig3-eps} shows a comparison of the GLFT method (orange dots) and the diagonalization approach (blue solid line) where the electron-perturber interactions are modeled via Omont's pseudopotential \cite{Omont}.
For this paradigmatic calculation we consider the $\Sigma$  potential energy curves relative to the $n=30$ electronic Rydberg manifold.
In the diagonalization approach the corresponding potential energy curves at $n=30$ are calculated using one manifold above $n=30$ and two additional manifolds below $n=30$.
On the qualitative level shown in \cref{fig:figs-fig3-eps}(a), both methods are in reasonable agreement over the whole range.
However, in the zoom-in plots in \cref{fig:figs-fig3-eps}(b) and (c) near the trilobite and butterfly molecular curves, respectively, the quantitative differences between the GLFT and diagonalization approaches are evident.
Note that the largest deviations are manifested near the avoided crossing between the trilobite and butterfly curve. Similar quantitative differences are seen in the quantum defect-shifted states (explored further in the following section) and, due to the extreme sensitivity of these weakly bound molecules, lead to noticeable disagreement between theoretical and experimental values for the binding energies.
Note that the largest deviations are manifested near the avoided crossing between the trilobite and butterfly curve. Similar quantitative differences are seen in the quantum defect-shifted states (explored further in the \cref{sec:resndisSr}) and, due to the extreme sensitivity of these weakly bound molecules, lead to noticeable disagreement between theoretical and experimental values for the binding energies.

In the butterfly molecular curves shown in \cref{fig:figs-fig3-eps}(c) the quantitative differences are more apparent, in particular, for internuclear distances to the left of the avoided crossing between the butterfly and the low-$l$ potential curve ``$32p$''.
The diagonalization method predicts that the wells in the butterfly curve are much shallower than those that are obtained within the GLFT framework.
For example, at distances $R=500~a_0$ the wells in the butterfly potential energy curve are $\sim30\%$ shallower than those of the GLFT approach whereas at shorter internuclear distances, i.e. $R\approx100-200~a_0$, the corresponding deviation between the two methods increases at $46\%$.
Such discrepancies in the relative depth of the wells strongly influences the number of vibrational states that they support.
These differences between the diagonalization treatment and more sophisticated methods are known \cite{Fey}, and stem from the lack of a systematic pathway to increase the accuracy of the diagonalization method. The convergence of the potential energy curves at given $n$ is not guaranteed by increasing the number of manifolds included in the basis.
However, the advantage of the GLFT approach is that its convergence does not suffer from these issues.
This occurs since the corresponding $K-$ matrix is evaluated using the energy-normalized Coulomb functions at the energy $\nu=\sqrt{-2 \epsilon(R)}$ of the electronic potential curves. 
The diagonalization method, on the other hand, uses the hydrogenic wave functions evaluated at hydrogenic energies. 
Thus, this approach requires hydrogenic wave functions from different Rydberg manifolds in order to minimize the errors of the potential energy curves at energy $\nu=\sqrt{-2 \epsilon(R)}$.

\begin{figure}[t!]
	\centering
	\includegraphics[scale=0.6]{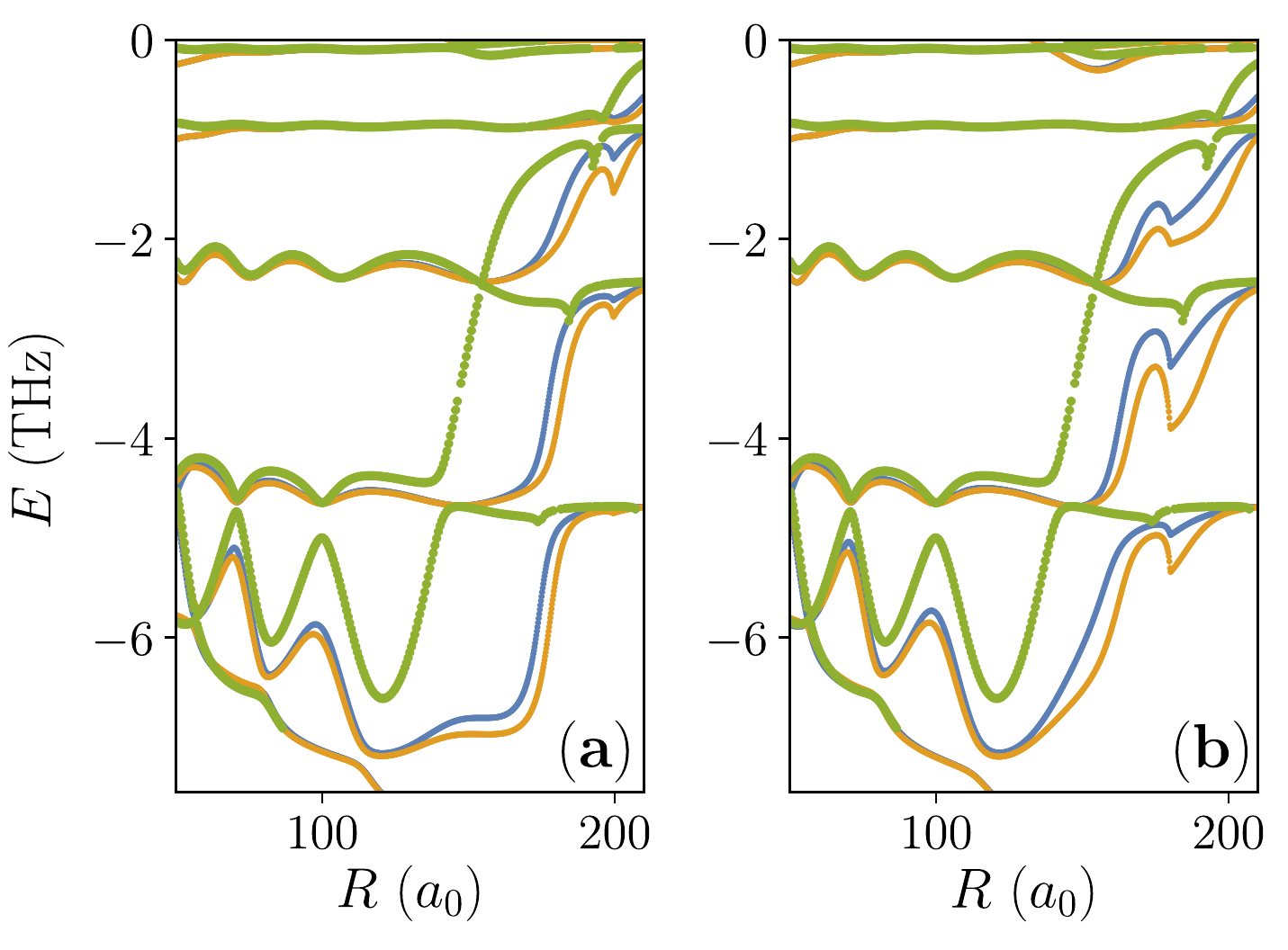}
	\caption{(color online) Comparison of the $\Sigma-$molecular potentials curves for $Rb^*-Rb$ at $n=10$ electronic Rydberg manifold. In panels (a) and (b) the potential energy curves obtained via GLFT method are indicated by the green points. The blue (orange) solid lines refer to the calculations within diagonalization framework where the corresponding results are converged by using one (two) manifold(s) above and two (three) below $n=10$. In addition for the diagonalization method the local momentum $k(R)$ is defined as $k(R)=\sqrt{2/R-1/n^2}$ for panel (a) whereas for panel (b) is set as $k(R)=\sqrt{2/R-1/(n-0.5)^2}$.}
	\label{fig:figs-fig4-eps}
\end{figure}

This issue becomes more severe at low $n$.  \cref{fig:figs-fig4-eps}(a) and (b) illustrate the $\Sigma$ potential energy curves of the $n=10$ Rb$^*-$Rb molecule. The differences between the GLFT approach (green dots) and the diagonalization method (blue and orange solid lines) are apparent even on a qualitative level.
In this figure we illustrate also the convergence challenges inherent in the diagonalization method; the blue solid line in \cref{fig:figs-fig4-eps}(a) and (b) is calculated by using the $n=8,9,10,11$ manifolds while the orange solid line is calculated using the $n = 7-12$ manifolds in the basis. 
This means that the calculations shown in orange use $50\%$ more basis states than the corresponding results shown in blue.
These particular choices of the truncated basis yield potential energy curves that are closest to the corresponding GLFT calculations, which use only 10 basis states.
It is apparent, though, that the results from the diagonalization method are still quite different both from the GLFT calculation and from each other.
Note, also, that the potential curves in orange shown in \cref{fig:figs-fig4-eps} use many more Rydberg manifolds than in the corresponding calculation shown in \cref{fig:figs-fig3-eps}, and in spite of this, the deviations from the GLFT results are starker.
Thus, as mentioned above, the inclusion of additional Rydberg manifolds in the diagonalization approach does not guarantee convergence, as was also shown in Ref. \cite{Fey}.

\begin{figure}[t!]
	\centering
	\includegraphics[scale=0.4]{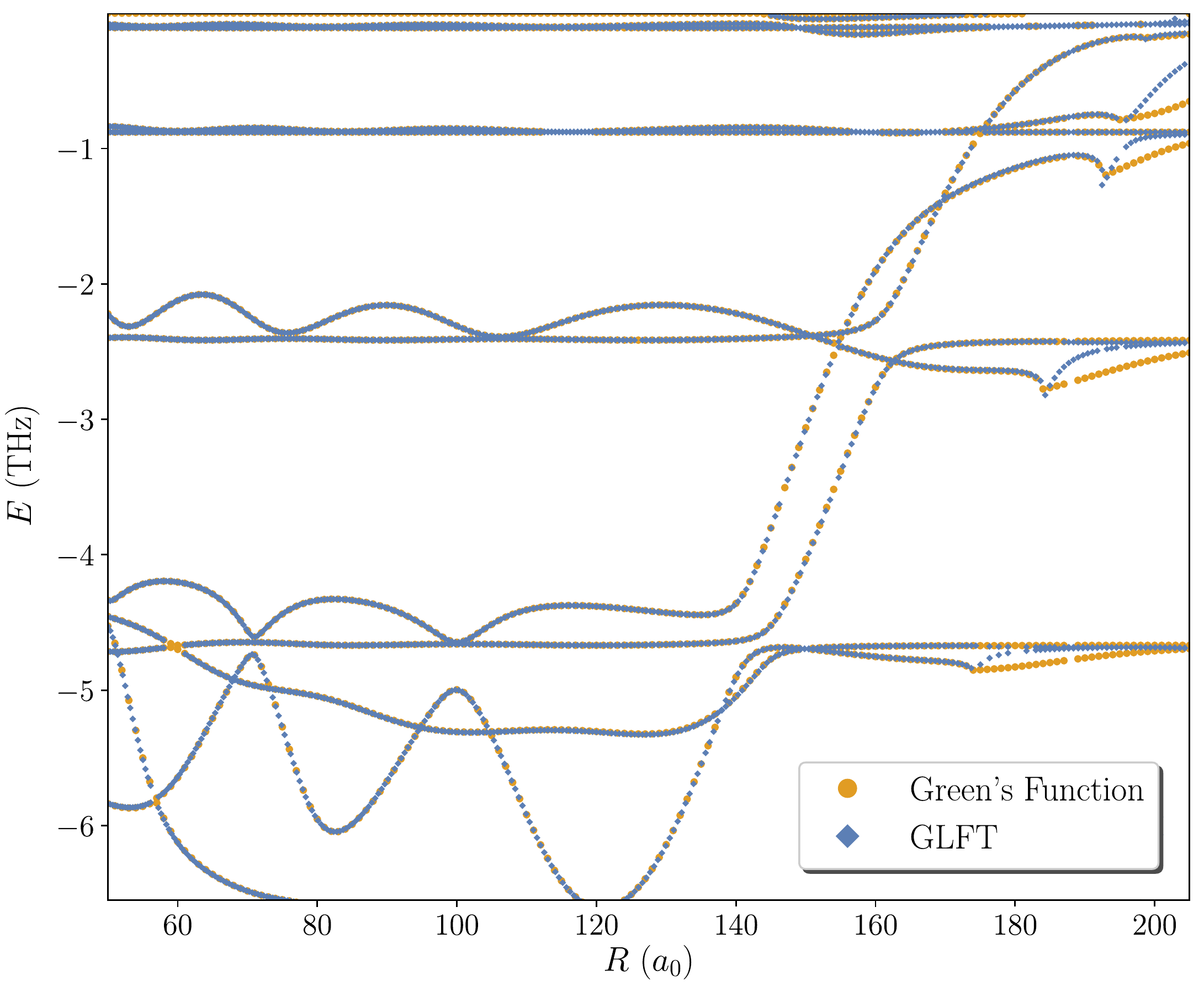}
	\caption{(color online) Comparison between the Green's function method (orange dots) and the GLFT framework (blue diamonds) for the $\Sigma-$ and $\Pi-$potential energy curves of $Rb^*-Rb$ at $n=10$ electronic Rydberg manifold.}
	\label{fig:figs-fig5-eps}
\end{figure}
In addition, in panel (a) the diagonalization approach depends on the local momentum $k(R)$ of the ''e-B'' subsystem which is defined relative to the $n=10$ manifold, i.e. $k(R)=\sqrt{2/R-1/10^2}$ whereas in panel (b) the local momentum is defined by the relation $k(R)=\sqrt{2/R-1/9.5^2}$.
Note that the local momentum in the diagonalization method is conventionally defined relative to the hydrogenic manifold, which is an ambiguous approximation resulting in an important disadvantage.
However, in the present method as well as in the Green's function approach the ''e-B'' momentum depends directly on the energy $\epsilon(R)$ of the potential curves, i. e. $k(R)=\sqrt{2/R+2\epsilon(R)}$, and thus it is determined in a self-consistent manner.

In \cref{fig:figs-fig4-eps} we observe that the butterfly potential curves obtained via the diagonalization method (see the blue and orange solid lines) are consistent with each other, however, they are inaccurate in predicting the range of internuclear distances $R$ where the $P-$wave resonance induces a steep drop in the energy of the butterfly potential curve.
For example, in \cref{fig:figs-fig4-eps}(a) we observe that the butterfly potential plunges down at larger $R$ for the diagonalization calculations than the corresponding ones in the GLFT method.
By changing the definition of the local momentum $k(R)$ in \cref{fig:figs-fig4-eps}(b) the results of the diagonalization approach shifted to internuclear distances closer the GLFT calculations but the discrepancies between the two methods are still apparent.

The robustness and the high level of accuracy of the present theory is illustrated in \cref{fig:figs-fig5-eps}, where
 we compare the results of the GLFT theory (blue diamonds) with those from the standard Green's function method (orange dots) \cite{hamilton_jpb_2002,khuskivadze_pra_2002}.
 Note that \cref{fig:figs-fig5-eps} depicts both the $\Sigma$ and the $\Pi$ Rb$_2$ potential energy curves. 
The agreement between the two methods is excellent over the entire range of internuclear distances $R$ up to the classical turning point, i.e. $R=2n^2$.
Some deviations are evident beyond the classical turning point; these stem from the fact that the Green's function method which we used is not designed to correctly address the regime where the momenta $k(R)$ becomes imaginary.
Computationally, the GLFT and the Green's function approach operate on the same level of efficiency.
However, we should stress that the Green's function method can be used only for bound state calculations in contrast to the GLFT approach which utilizes the concept of the $K-$matrix, enabling the treatment of both continuum and bound states without compromising its computational efficiency.

\subsection{Strontium}\label{sec:resndisSr}
Our discussion of the potential energy curves of Rb$^*+$Rb Rydberg molecules emphasized their most prominent features, namely the potentials associated with the trilobite and butterfly states.
 These potentials, as a result of the high degeneracy of states in the Rydberg manifold, are typically many GHz deep. 
The remaining potential energy curves, associated with the non-degenerate quantum defect-shifted Rydberg states with low angular momentum, are the subject of the present section.   
The typical well depths are $10-100$ MHz.  
The potential energy curve associated with a non-degenerate Rydberg state closely resembles the radial electronic wave function modulated by the $S$-wave and $P$-wave interactions with the perturber.
Since we have confirmed in the previous section that the GLFT method circumvents the pitfalls inherent in attempts to increase the accuracy of the Fermi/Omont pseudopotentials and gives converged results, we now apply it to the calculation of the potential energy curves of Sr$(ns)$+Sr Rydberg molecules.
These are ideal to study with the GLFT method because the commonly studied isotope, $^{84}$Sr, does not have a hyperfine structure, the $ns$ Rydberg states have no fine structure, and the spin-orbit splitting of the $P$-wave phase shift is expected to be small.
As a result, the spin-independent theory developed here can be applied immediately, although we note in passing that the modular nature of the GLFT method can be readily extended to a spin-dependent Hamiltonian in future applications.
By comparing the GLFT results with experimental signatures of these strontium Rydberg molecules, we can further benchmark the accuracy of the GLFT theory.

The vibrational spectra of these molecules are reported in Refs.~\cite{killian2015,killian2016} for principal quantum numbers ranging from $n = 29$ to $n=38$.
The $n$-dependence of the binding energies, $E\sim(n-\tilde\mu_s)^{-6}$, follows the anticipated Rydberg scaling law. 
Each potential curve supports around four vibrational states over this range of principle quantum numbers. 
These measurements form the basis for a series of experiments at increasingly high density and principal quantum number. 
In such regimes many perturbers can be located within the Rydberg orbit, and the experimental spectra exhibit features at energies which are sums of two or more dimer energies and are thus associated with polyatomic states of two, three, and more Sr perturbers \cite{Whalen2,WhalenPoly}. 
At even higher densities and Rydberg levels the observed line shapes reveal details of the many-body response of the gas to the Rydberg impurity.
Since the properties of a Rydberg atom embedded in these more exotic dense perturber scenarios are fundamentally linked to the energy levels of the strontium dimer, the GLFT calculations presented here serve a second purpose: to better calibrate these potential energy curves and the underlying atom-electron phase shifts. 

As stated at the beginning of \cref{sec:resndisRb}, to obtain potential curves for a new atomic species we only need to modify the quantum defects and scattering phase shifts which are the inputs to the GLFT calculation. The quantum defects of Sr are summarized in Ref.~\cite{eiles_revJPB_2019a} and the electron-atom scattering phase shifts for Sr were calculated in Ref.~\cite{bartschat_jpb_2003}. 
It is clear from our GLFT calculations that these phase shifts are not accurate enough to obtain a theoretical spectrum in agreement with experiment. 
At large internuclear distances the calculated potential curves are too deep: the ground state molecules are more deeply bound in the calculation than observed in the experiment.
At smaller internuclear distances the potential curves are insufficiently repulsive.
The calculated excited states, particularly the 2-3 most highly excited ones, leak into the short-range region of strong ion-atom attraction, and their energies do not match the corresponding experimental measurements.
It is not surprising that the calculated phase shifts at these very low scattering energies require tuning in order to obtain quantitative agreement (on the level of a few MHz at low $n$ to even a few hundred kHz at higher $n$); this has been necessary in Rb as well.
Rather than attempting a new fitting of the phase shifts, we adopt the same parameterization as in the theoretical analysis of Ref.~\cite{killian2015}.
There, the effective range theory for the energy-dependent $S$-wave scattering length $a_S(k)$ and a constant parameter for the energy-dependent $P$-wave scattering volume $a_P^3(k)$ was used:
\be
\label{eq:scparms}
a_S (k)= a_S(0) + \frac{\pi \alpha}{3}k\,\,,\,\,a_P^3(k) = \left[a_P(0)\right]^3.
\ee
The two fit parameters $a_S(0)$ and $a_P(0)$ are modified until the predicted vibrational spectra matches the measured spectra. 
Although this is a straightforward way to parametrize the energy-dependent phase shifts, we note that it has two key limitations. 
While $a_S(k)$ is the valid expansion of the scattering length in the $k\to0$ limit, as $k$ increases $a_S(k)$ in \cref{eq:scparms} rapidly becomes only qualitatively accurate due to the presence of additional $k^2$ and $k^2\ln k$ terms in the expansion.
This means that although it can be used to obtain the zero-energy scattering length to a reasonable degree of confidence, it is not quantitatively reliable at smaller internuclear distances. This will be reflected in the binding energies of excited states probing these distances.

On the other hand, a constant scattering volume is unphysical as $k\to 0$, since the threshold law of the phase shift is quadratic in k as can be seen in Born approximation.
This is typically not problematic in the calculation of Rydberg molecule potential curves since the $P$-wave contribution is much smaller than the $S$-wave interaction at distances $R$ where this unphysical nature of $a_P$ is most prominent, i.e. in the vicinity of the classical turning point ($R\approx2n^2$).
Although this parametrization could in principle be at least qualitatively accurate at larger $k$ values, we note that the phase shifts of Ref.~\cite{bartschat_jpb_2003} exhibit significant energy dependence and this is only a crude approximation. 
We therefore emphasize that both of these parametrizations, but especially the effective $P$-wave scattering volume, are only convenient parametrizations for fitting and imply only very generic properties of the phase shifts except for the zero-energy scattering length, which can be fit quite accurately using only the ground vibrational state's binding energy.
We obtain the phase shifts from \cref{eq:scparms} using
\be
\label{eq:scatlengths}
\delta_S(k) = -\tan^{-1}\left[ka_S(k)\right],\,\,\delta_P(k) = -\tan^{-1}\left[k^3a_P^3(k)\right].
\ee

\begin{table}[t!]
	\centering
\begin{tabular}{ccccccc}
& GLFT & PT & $a_S(0)$ & $a_P(0)$ & polarization potential& \\
\hline
(i) &  \ding{51} & \ding{55} & -12.65 & 9.6 &  \ding{51}&\\
(ii) &  \ding{55} &  \ding{51} & -13.2 & 8.4 &  \ding{55} &\\
(iii)&\multicolumn{6}{c}{ measurements of Ref.\cite{killian2015}\,\,\,\,\,\,\,}{}\\
(iv) &  \ding{55} &  \ding{51} & -13.2 & 9.8 &  \ding{51} &\\
(v) &  \ding{51} & \ding{55} & -13.2 & 8.4 &  \ding{55} &
\end{tabular}
	\caption{ Definitions of the model calculations described in the text. PT refers to the potential (\cref{eq:ptpot}) obtained within first-order perturbation theory; polarization potential refers to the ion-atom potential $-\alpha/2R^4$. }
	\label{tab:tab1}
\end{table}

\begin{figure}[t!]
	\centering
	\includegraphics[width=\columnwidth]{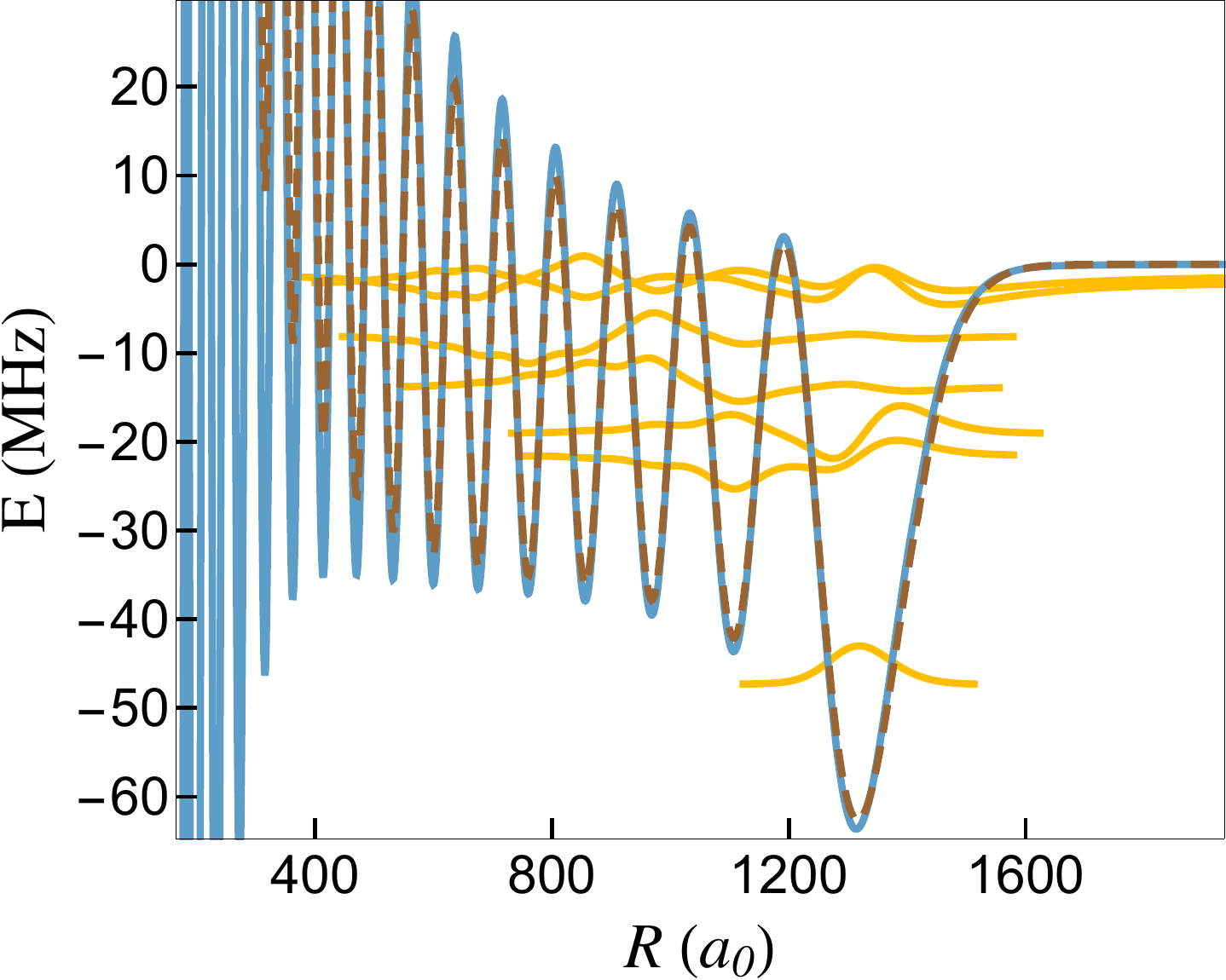}
	\caption{(color online) Comparison of the potential energy curves from model calculations (i) (blue solid) and (ii) (brown dashed) for the Sr(30s)+Sr Rydberg dimer. These models are defined in the text.  The dissociation threshold at zero energy is the quantum defect-shifted energy of the 30s state. The yellow curves show the vibrational bound states calculated in the model (i), defined in the text and in \cref{tab:tab1}.}
	\label{fig:figs-fig6-eps}
\end{figure}

\cref{fig:figs-fig6-eps} depicts two potential energy curves for the Sr(30s)+Sr Rydberg molecule. In the solid blue color we show the potential energy curve calculated using the GLFT approach  with $a_S(0) = -12.65~a_0$ and $a_P(0) = 9.6~a_0$. 
We refer to this model calculation as (i); within each model we calculate a family of potential curves for each $n$ using these same parameters and computational technique. 
Importantly, (i) also includes the ion-atom polarization interaction, $-\frac{\alpha}{2R^4}$, where $\alpha\approx 186$~a.u. is the polarization of Sr. 
This attractive interaction is strongest at distances $R<500~a_0$ and was neglected in Ref.~\cite{killian2015}. 
The dashed brown curve in \cref{fig:figs-fig6-eps} shows the potential energy curve used in Ref.~\cite{killian2015}, which employed a model potential from first-order perturbation theory (PT) to treat $S$ and $P$-wave electron-perturber scattering,
\be
\label{eq:ptpot}
V_\text{PT}(R) = 2\pi a_S(k_s)\left|\Psi_{s}(R)\right|^2 + 6\pi a_P^3(k_s) \left|\frac{d\Psi_{s}(R)}{dR}\right|^2.
\ee
The semiclassical electron momentum is $k_s = \sqrt{2/R-1/(n-\tilde\mu_s)^2}$ and $\Psi_s(R)$ is the Rydberg $ns$ wave function.   We refer to the model calculation using \cref{eq:ptpot} with $a_S(0) = -13.2~a_0$ and $a_P(0) = 8.4~a_0$ as (ii). 
The $a_S(0)$ value in each case was obtained by fitting the binding energy of the ground vibrational state to the measured value. 
Since the $P$-wave interaction and ion-atom interaction are very weak at such large $R$ values this state's energy is essentially fixed by the strength of the $S$-wave interaction. 
After fitting $a_S(0)$, $a_P(0)$ was fit by aligning the theoretical and experimental binding energies for the higher excited states. 
The vibrational wave functions bound in potential (i) are shown also in \cref{fig:figs-fig6-eps} in yellow. 
Those obtained using potential (ii), not shown, are nearly identical and have very similar binding energies, despite the differences in the potential energy curves. 
Since the $P$-wave interaction in both models is repulsive enough to restrict nearly all of the vibrational states to distances larger than $500~a_0$, the polarization potential has only a small effect. 
In our numerical calculations for the vibrational states a hard-wall cut-off is imposed for the atom-ion interaction at distances $R\sim 100 -200~a_0$. Our results are insensitive to this cut-off since the vibrational states are mostly localized in the outer well of the potential curves (see \cref{fig:figs-fig6-eps}).

\begin{figure}[t!]
	\centering
	\includegraphics[width=\columnwidth]{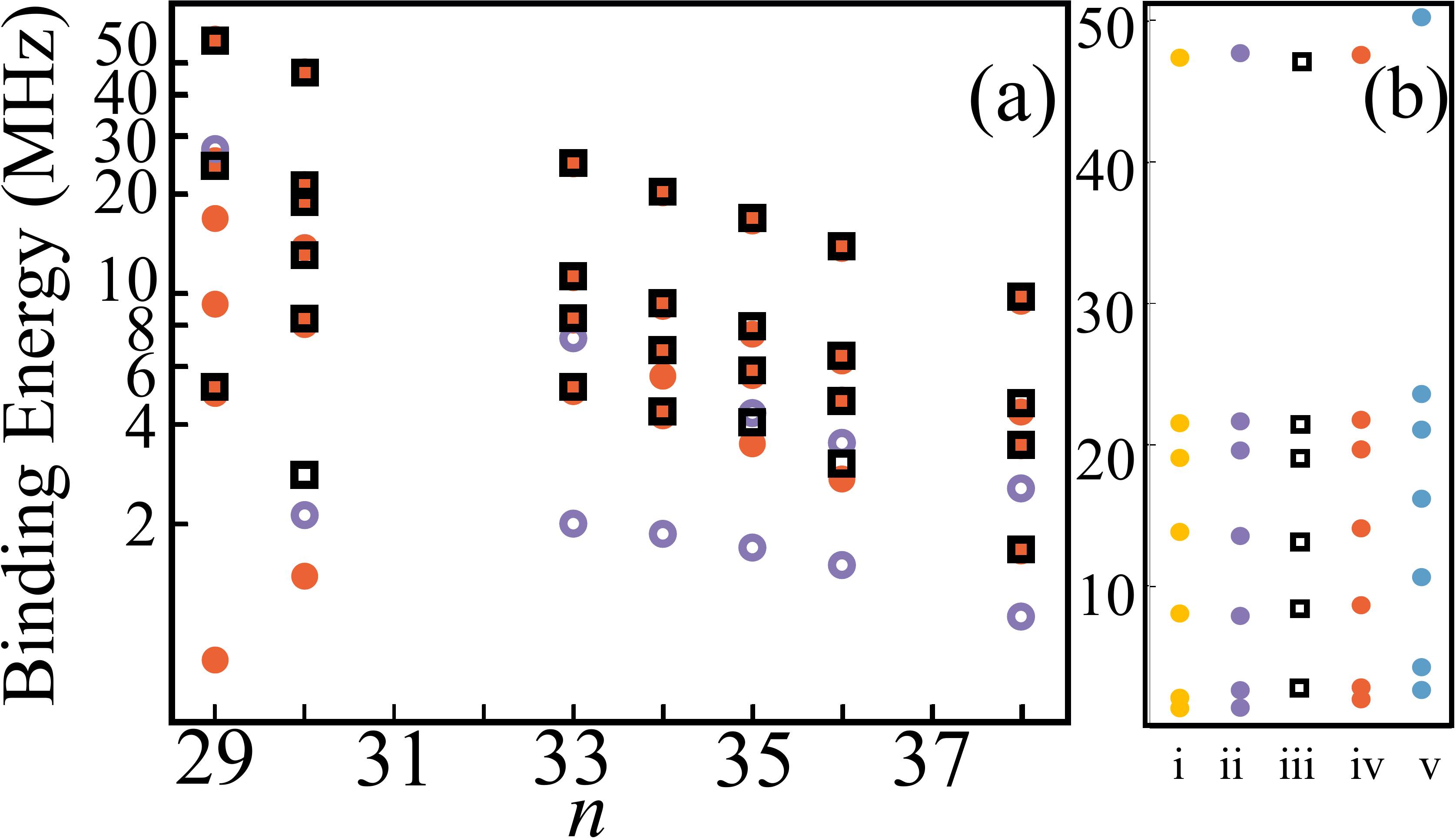}
	\caption{(color online) (a) Comparison of calculated (circles) and measured (squares) spectra for strontium Rydberg molecules.
The theoretical calculations are based on model (i), defined in the text and \cref{tab:tab1}.
The vibrational states whose binding energies are marked with open purple circles have substantially weaker line strengths than those marked by filled orange circles. Experimental data for the missing $n$ states are not available. (b) Comparison of calculated binding energies for the $30s$ Rydberg molecule using the four models defined in \cref{tab:tab1}: (i), (ii), (iv), and (v) and the experimental values (iii).  Note that a log-linear plot is used in (a) and a linear plot in (b).}
	\label{fig:figs-fig7-eps}
\end{figure}

\cref{fig:figs-fig7-eps}(a) compares the binding energies computed using the GLFT approach (model (i)) with the measurements reported in Ref.~\cite{killian2015}. 
This demonstrates that the fitted $a_S(0)$ and $a_P(0)$ values used in model (i) are sufficient to closely reproduce the measured spectra over this whole range of $n$ values. 
To verify that the fit parameters are independent of $n$ we chose to fit them to the vibrational energies using only the $30s$ spectrum. 
The filled orange circles mark states whose line strengths are roughly equal, while the open purple circles mark states whose line strengths are less than 20\% as strong as the average orange point.  
The line strengths are computed using a simple approximation for the Franck-Condon factor assuming a flat initial scattering state between the two atoms, as in Ref.~\cite{killian2015}. 

\cref{fig:figs-fig7-eps}(b) shows the results of vibrational states using the different models which are summarized in \cref{tab:tab1}.
More specifically, in \cref{fig:figs-fig7-eps}(b) we compare only the $30s$ data, but now also for a perturbative calculation using \cref{eq:ptpot} including the polarization potential.
 We label this calculation as model (iv).
By this additional comparison we are able to better understand the way in which the fit parameters compensate for inaccuracies in the $30s$ potential curves, whether these stem from the choice of method (GLFT or PT) or from the inclusion of the polarization potential. 
To obtain a theoretical spectrum with model (iv) matching the experimental data when including the polarization potential, we had to re-fit the $a_P(0)$ parameter to $a_P(0) = 9.8~a_0$. 
Although the model calculation (ii) used in Ref.~\cite{killian2015} apparently ignored the polarization potential, this comparison shows that this error was compensated for by the difference $\Delta a_P = 1.4~a_0$ between models (ii) and model (iv). 
Fig. 7(b) shows also the results of  model (v) which refers to a GLFT calculation using the scattering parameters of model (ii) and neglecting the polarization potential. 
This method clearly fails to predict the measured binding energies. 

\subsubsection{Lessons learned from different models}
From these calculations, we arrive at three conclusions. 
First, that the GLFT model using fitted effective electron-atom scattering phase shifts (model (i)) reproduces, for all $n$ values where measurements exist, the experimental spectra.
Complementing the theoretical comparisons in \cref{sec:resndisRb}, this provides an experimental confirmation of the validity of the GLFT approach. 
Some minor discrepancies are, however, visible in the comparison. 
The theory predicts weakly bound vibrational states (binding energies of 3MHz or less) with weak but finite line strength which are not observed in the experiment. 
These might be suppressed in the experiment by additional contributions to the line strength not accounted for in our approximate Franck-Condon factors, or they could be obscured by the atomic resonance, or indeed they could signal an additional energy dependence that cannot be compensated for by fitting a constant $P$-wave scattering volume or which stems from the higher-order effective range terms neglected in $a_S(k)$. 
A more glaring discrepancy is that the theory predicts two deep bound states for the $29s$ state with strong line strengths which are not observed. 
 The theory calculations of Ref.~\cite{killian2015} using model (ii) show also these extra vibrational bound states; based on the trends of the other vibrational states as a function of $n$ one would expect these states to exist. 
Further study of the experimental spectrum is necessary to resolve this issue and identify if it has an origin in the experimental setup or if it heralds additional physics not included in the theory. 

Our second conclusion is that the electron-strontium scattering length must be in the neighborhood of $-12.65~a_0$. 
We estimate an uncertainty of 0.1 atomic units on this value based on the fit of $a_S(0)$. 
The extracted scattering length from model (ii), -13.2 $a_0$, is about 5\% different; this difference can be attributed to the known differences between calculations using the Fermi pseudopotential and the GLFT calculation. 
Indeed, in the trilobite potentials dominated by $S$-wave scattering presented in our study of Rb (see \cref{fig:figs-fig3-eps}(b)) differences on this order were already visible.
The $a_S(0)$ values obtained using the models (i) and (ii) both differ quite strongly, by 25\%, from the calculated scattering length \cite{bartschat_jpb_2003}. 
Although the authors of Ref.~\cite{killian2015} attributed this to the uncertainty in the potential energy calculation, we can now claim on more rigorous theoretical grounds that the greater part of this difference is due to a real discrepancy between the scattering length provided in Ref.~\cite{bartschat_jpb_2003}, and only the 5\% error between $-12.65~a_0$ and -13.2~$a_0$  is due to the approximations made in the perturbative potential energy curve of model (i). 
That the calculated scattering length is overestimated is not surprising due to the challenges of converging the atom-electron scattering calculations at such low energies. 

Third, we conclude that the $P$-wave scattering volume must be large and positive in order to produce a repulsive barrier at short-range which suppresses the effect of the polarization potential and localizes the vibrational states in the outer wells. 
This implies that the $P$-wave phase shift must be negative and significantly larger in magnitude than given by the calculations, particularly at high $k$ values. 
These results show that further work is necessary in order to produce a more accurate set of scattering phase shifts.
Caution is also warranted since, as seen in \cref{fig:figs-fig6-eps,fig:figs-fig7-eps}, the scattering parameters and even the potential energy curves themselves are not uniquely determined by the binding energies, and fitting of the binding energies alone is likely insufficient without additional theoretical calculations of the strontium-electron phase shifts to pin down these phase shifts.

\section{Summary and Conclusions}\label{sec:sumncon}

\subsection{Summary of the method}
\begin{figure}[t!]
	\centering
	\includegraphics[scale=0.45]{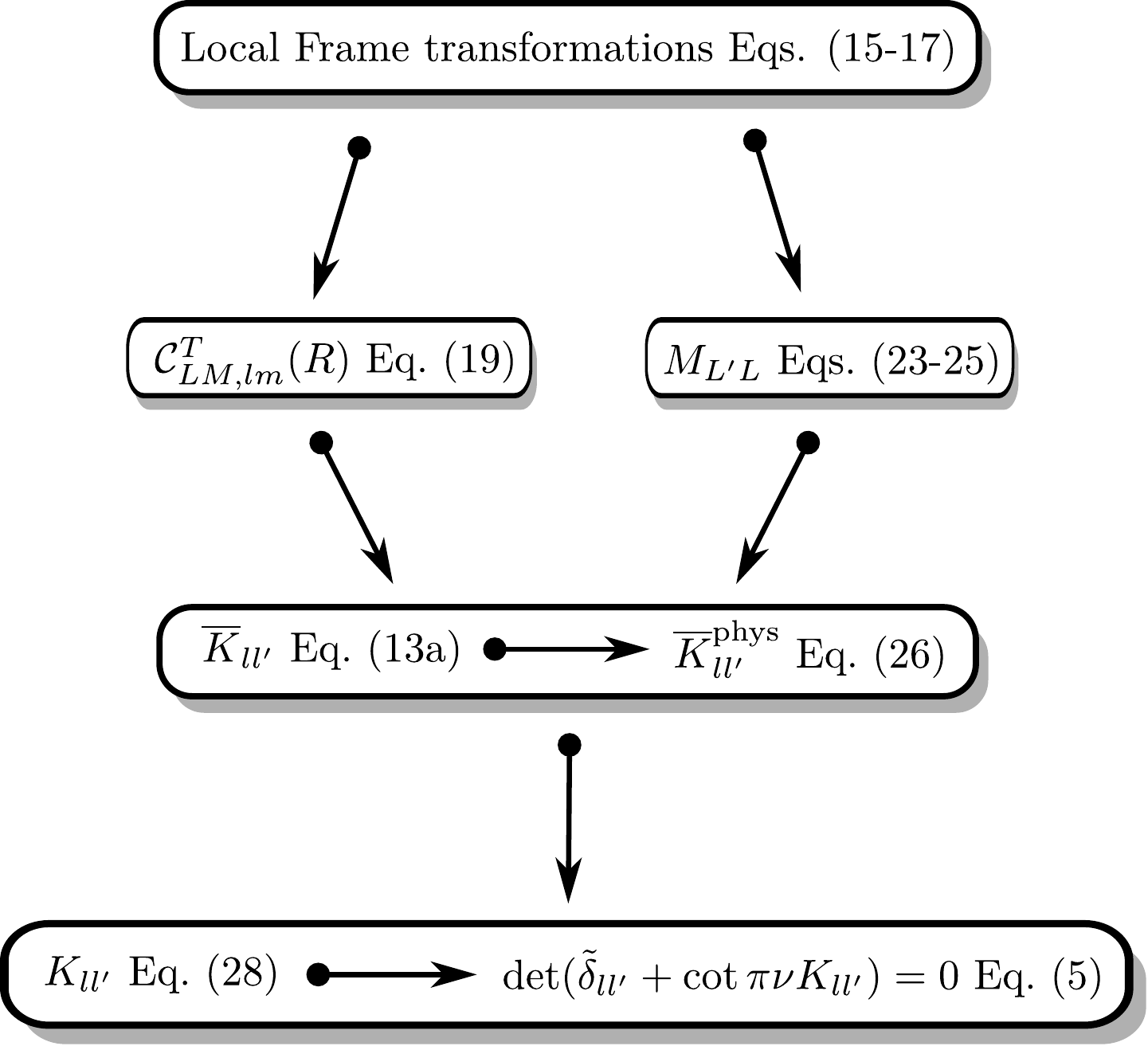}
	\caption{(color online) An outline of the key expressions utilized in the GLFT method.}
	\label{fig:figs-fig8-eps}
\end{figure}
Since the derivation of the GLFT approach for Rydberg molecules involves many steps, we provide here a summary of the key equations that must be implemented in order to utilize this method. 
Also, note that \cref{fig:figs-fig8-eps} outlines the summary of these key expressions.
First, \cref{eq:kmatrix1,eq:kmatrix1b,eq:kmatrix1c} define the $\overline{K}$ matrix which is associated with the QDS energy-normalized regular and irregular Coulomb functions.
Using the Schwinger identity, the $\overline{K}$ matrix acquires a separable form and depends on two terms: the $\mathcal{C}^T_{LM,lm}(R)$ and the $M_{L'L}$ matrix elements.
These are expressed in terms of the local frame transformation in \cref{eq:cterms} and in \cref{eq:mmatlft,eq:g0gcterm,eq:lambdalft}, respectively.
\cref{eq:lft} provides the local frame transformation relation which connects the QDS Coulomb functions centered at the Rydberg core with the regular field-free functions around the perturber.

In the spirit of MQDT theory, the strongly closed channels of the $\overline{K}$ matrix are eliminated.
This elimination step yields the physical $\overline{K}$ matrix, namely $\overline{K}^{\rm{phys}}$, which is given in \cref{eq:physkbar}.
In a final step, the physical $\overline{K}^{\rm{phys}}$ is connected to the $K$-matrix via the expression  \cref{eq:kkbar}.
Note that $K-$matrix is associated with the energy-normalized pair of regular and irregular Coulomb functions.

For a specific Rydberg molecule two inputs are required: the atomic quantum defects $\mu_l$ and the electron-atom scattering phase shifts $\delta_L$.
Using these two parameters, the $K$-matrix is inserted in the determinantal equation \cref{eq:determinantal1}.
The roots $\nu(R)$ of \cref{eq:determinantal1}  determine the molecular energy curves via $\epsilon(R)=-\frac{1}{2\nu^2}$.
\subsection{Conclusions}
Based on the generalized local frame transformation theory we have developed a formalism to describe asymmetrically excited three-body systems, exemplified here with an excited Rydberg atom interacting with a ground state atom.
By employing the key concept of the local frame transformation, we have obtained closed form analytical formulas for the body-frame $K-$matrix associated with a diatomic ultra-long-range Rydberg molecule.
The potential energy curves are then obtained by numerically solving the one-dimensional determinantal equation expressed in terms of the $K$-matrix.
We have shown that this GLFT approach provides potential energy curves which are known to be more accurate than those obtained via the diagonalization approach, which cannot be rigorously converged.
We have used this advantage of the GLFT to re-analyze the vibrational spectra of strontium Rydberg molecules, and have seen that it can be used to extract a more accurate zero energy scattering length from experimental measurements.
One major advantage of the GLFT method over the Green's function treatment, which operates at the same level of accuracy, is that it can be easily extended to other physical systems, for example to Rydberg atoms in the presence of multiple perturbers.
This would provide quantitative improvements to the theory of polyatomic Rydberg molecules and Rydberg composites developed in \cite{WhalenPoly,eiles2016ultracold,hunter2019rydberg}.
In addition, due the modularity of the GLFT toolkit, the local frame transformation for Rydberg molecules presented here can be combined other frame transformations in order to investigate more complicated physical systems, e.g. Rydberg molecules in external electric fields or include relativistic effects.
Another possible and less straightforward application of the present theory is related to the case where a molecule is used as a perturber \cite{rittenhouseUltracoldGiantPolyatomic2010,gonzalez-ferezRotationalHybridizationControl2015a}.
In this particular case, the frame transformation derived here can be extended in order to incorporate the charge-dipole interaction which influences the motion of the Rydberg electron, especially if the molecule is a polar one.
The resulted frame transformation can be combined with the LFT theory of Clark in Ref.\cite{clarkElectronScatteringDiatomic1979} expressing the electronic wave function from the body-frame of the Rydberg molecule to the body-frame of the molecular perturber.
This step permits us to connect to the electron-molecule $K-$matrix in an analogous way as was shown here. 

\begin{acknowledgments}
We thank Chris H. Greene and Hossein R. Sadeghpour for discussions.
MTE is grateful to the Alexander von Humboldt stiftung for support through a postdoctoral fellowship.
FR acknowledges the financial support by the U.S. Department of Energy (DOE), Office of Science, Basic Energy Sciences (BES) under Award No. DE-SC0012193.
The numerical calculations have been performed using NSF XSEDE Resource Allocation No. TG-PHY150003.
PG acknowledges E. Diamantopoulos for helpful discussions at the early stages of this work.
\end{acknowledgments}
\bibliography{references.bib}
\end{document}